\documentclass{PoS}

\title{Poisson Statistics in the High Temperature QCD Dirac Spectrum}

\ShortTitle{Poisson Statistics in the High-T QCD Dirac Spectrum}

\author{\speaker{Tam\'as G.\ Kov\'acs}\thanks{Supported by OTKA Hungarian
    Science Fund grant 49652 and EU Grant (FP7/2007-2013)/ERC N$^o$208740.}  \\
        Department of Physics, University of P\'ecs \\
        H-7624 P\'ecs, Ifj\'us\'ag \'utja 6, Hungary \\
        E-mail: \email{kgt@fizika.ttk.pte.hu}}

 \author{Ferenc Pittler \\
        Department of Physics, University of P\'ecs \\
        H-7624 P\'ecs, Ifj\'us\'ag u. 6, Hungary \\
        E-mail: \email{pittlerferenc@gmail.com}
}

\abstract{At low temperature in the epsilon regime of QCD the low-end of the
  Dirac spectrum is described by random matrix theory. In contrast, there has
  been no similarly well established staistical description in the high
  temperature, chirally symmetric phase. Using lattice simulations we show
  that at high temperature a band of extremely localized eigenmodes appear at
  the low-end of the Dirac spectrum. The corresponding eigenvalues are
  statistically independent and obey a generalized Poisson
  distribution. Higher up in the spectrum the Poisson distribution rapidly
  crosses over into the bulk distribution predicted by the random matrix
  ensemble with the corresponding symmetry. Our results are based on quenched
  lattice simulations with the overlap and the staggered Dirac operator done
  well above the critical temperature at several volumes and values of
  $N_t$. We also discuss the crucial role played by the fermionic boundary
  condition and the Polyakov-loop in this phenomenon.}

\FullConference{The XXVIII International Symposium on Lattice 
                Field Theory, Lattice2010 \\
		June 14-19, 2010 \\
		Villasimius, Italy}

\begin{document}

\section{Introduction}

The lowest part of the spectrum of the QCD Dirac operator encodes important
physical information concerning the low energy behavior of strongly
interacting systems.  In the so called epsilon regime of the low temperature
phase the statistics of the lowest Dirac operator eigenvalues is described by
chiral random matrix theory (RMT). This is a well established fact supported
by analytic calculations in the corresponding low energy sigma model as well
as extensive numerical lattice studies (see e.g.\ \cite{Verbaarschot:2000dy}
for a summary of results and a list of original references). In contrast,
above the finite temperature transition ($T_c$) there is no generally accepted
statistical description of the low end of the Dirac spectrum. Since in this
regime there is no first principles analytic information available, to a first
approximation the Dirac operator here can be regarded as a large fluctuating
random matrix, with its size going to infinity in the thermodynamic limit.

From this perspective there are two possible types of fundamentally different
extreme behavior the statistics of the lowest part of the spectrum can follow.
If typical fluctuations cannot mix eigenmodes nearby in the spectrum, 
eigenmodes have to be localized, the corresponding eigenvalues are expected to
be statistically independent and follow essentially Poisson statistics. If on
the other hand, nearby eigenmodes in the spectrum can easily mix, they
necessarily become delocalized and the spectrum is expected to obey random
matrix statistics. Lattice simulations can decide which scenario is realized
in nature.

Above $T_c$ chiral symmetry is restored and the density of eigenvalues of $D$,
the order parameter of chiral symmetry breaking, vanishes. Random matrix
theory has specific predictions for the eigenvalue statistics around such
a ``soft edge'' \cite{Forrester}. Lattice simulations, however, failed to
reproduce the RMT predictions for the spectral statistics at
the spectrum edge \cite{Farchioni:1999ws,Damgaard:2000cx}. On the other hand,
bulk random matrix statistics for full Dirac spectra above $T_c$ were verified
previously \cite{Pullirsch:1998ke}. Based on lattice simulations around the
critical temperature, Ref.\ \cite{GarciaGarcia:2006gr} suggested that
around the chiral transition at $T_c$ a gradual change of eigenvalue
statistics at the edge occurs from RMT towards Poisson. Very recently, however,
Ref.\ \cite{Gavai:2008xe} argued that although low Dirac eigenmodes are
localized, localization appears to be a finite volume artifact disappearing in
the thermodynamic limit. If true, this would suggest RMT statistics for
eigenvalues at the spectrum edge. Lattice results obtained so far are thus
rather controversial.

In the present paper we provide some explanations for these apparent
controversies and draw a clear picture of the eigenvalue statistics above
$T_c$. The new ingredient in our analysis is that we study the eigenvalue
statistics separately in different regions of the spectrum starting with the
lowest eigenvalues and going upwards. Our main result is that the lowest part
of the spectrum consists of localized, independent eigenmodes obeying Poisson
statistics. Eigenmodes higher up in the spectrum gradually become more
delocalized and at the same time the eigenvalue statistics crosses over to
the bulk random matrix statistics that was previously seen in lattice
simulations. The phenomenon we report here is analogous to Anderson
localization in conducting crystalline solids with disorder. In that case
disorder can render electronic states at the band edge localized and
non-conducting while states deep in the band can still remain conducting and
delocalized. 

\section{Simulation details}

We performed simulations of the quenched $SU(2)$ gauge theory at a temperature
well above the critical temperature. Here the Polyakov loop $Z(2)$ symmetry is
spontaneously broken and for the quenched theory the two $Z(2)$ sectors are
equivalent. In contrast, the Dirac spectrum is known to depend strongly on the
Polyakov loop sector through the lowest Matsubara mode
\cite{Bilgici:2009tx}. It is in fact the combination of the Polyakov sector and
the antiperiodic quark boundary condition (b.c.) in the time direction that
determines the lowest part of the Dirac spectrum. If this combination
corresponds effectively to periodic b.c.\ (no twist), the spectral density
does not vanish at zero. If on the other hand, the combined effective b.c.\ is
antiperiodic (a twist of $-1$), the spectral density vanishes at zero and
there might even be a gap in the spectrum there.

In the presence of light dynamical quarks with antiperiodic b.c.\ the small
modes in the twisted sector would suppress the determinant and only the
positive Polyakov loop sector survives for large volumes. This is how the
fermion action breaks the Polyakov loop $Z(2)$ symmetry above $T_c$
\cite{Kovacs:2008sc}. For this reason we used for our study only
configurations in the positive Polyakov loop sector to mimic the most
important effect of dynamical quarks on the Dirac spectrum.

We analyzed spectra of both the staggered and the overlap Dirac operator on
$N_t=4$ and $N_t=6$ configurations with spatial volumes between $12^3-48^3$,
all at the same physical temperature $T=2.6T_c$. The configurations were
generated with the Wilson gauge action at $\beta=2.60$ and $2.725$. The
staggered and the overlap Dirac operator gave qualitatively similar results
and both data sets support our main findings. More details of this study can
be found in \cite{Kovacs:2009zj} for the overlap and in \cite{Kovacs:2010wx}
for the staggered Dirac operator.  In the following we shall present some
results with both types of lattice Dirac operator.

\section{Localization of small modes}

At first we directly measured the localization of the eigenmodes in different
regions of the low end of the spectrum. Instead of the most commonly used
quantity, the inverse participation ratio (IPR), we used the quantity
\begin{equation}
 {\cal V}_\psi = \left[ \sum_x (\psi^\dagger \psi(x))^2 \right]^{-1}
\end{equation}
for characterizing the localization of the normalized eigenmode $\psi$. ${\cal
  V}_\psi$ can be thought of as an approximate measure of the four-volume
occupied by the eigenmode. This can be seen by considering an idealized
eigenmode that is constant in a given subvolume $v$ of the total volume $V$
and zero elsewhere. We assume that at very high temperatures the eigenmodes
are maximally spread in the (short) time direction and define a length scale
\begin{equation}
 d_\psi = \left[ \frac{{\cal V}_\psi}{N_t} \right]^{1/3}
\end{equation} 
measuring the linear spatial extension of the eigenmode $\psi$. 

\begin{figure}
%\vspace{1cm}
\begin{center}
\includegraphics[width=0.6\columnwidth,keepaspectratio]{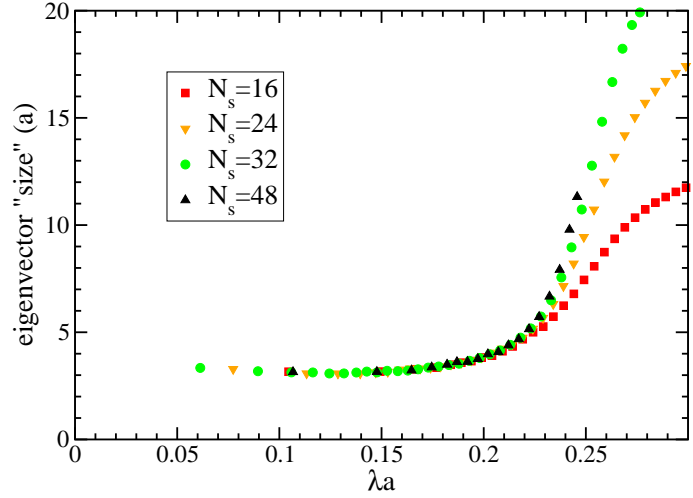}
\caption{\label{fig:em_size} The average linear extension (in lattice units)
  of staggered eigenvectors, $d_\psi$, as a function of the corresponding
  eigenvalues. The different symbols correspond to spatial box sizes
  $N_s=16,24,32,48$. All the ensembles have $N_t=4$.}
\end{center}
\end{figure}

In Fig.\ \ref{fig:em_size} we plot how the average linear size of eigenmodes
change as we go upwards in the spectrum starting from the lowest modes. It is
apparent that in the lowest part of the spectrum the eigenmodes are very
localized and their size is independent of the box size. Higher up in the
spectrum the eigenmodes spread out and they start to be constrained by the
finite box. There is a rather sharp transition point between the two types of
behavior, analogous to the mobility edge separating conducting and
non-conducting electron states in disordered conductors. 

\begin{figure}
\begin{center}
\includegraphics[width=0.6\columnwidth,keepaspectratio]{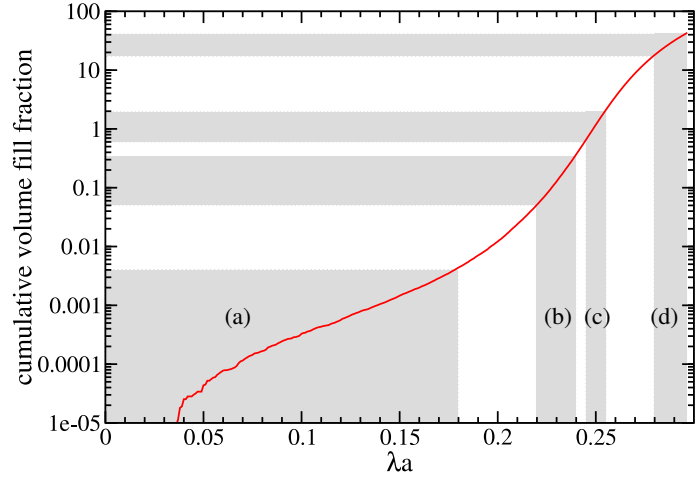}
 \caption{\label{fig:evsvolume24} The cumulative volume fill fraction for the
   $24^3\times 4$ staggered ensemble. The unfolded level spacing distribution
   will be computed separately in the four shaded spectral regions marked by
   (a)-(d) (see Fig.\ 4).} %\ref{fig:ks_levsp}).}
\end{center}
  \vspace{-0.5cm}
\end{figure}

Besides spatial localization, the other quantity affecting the mixing of
nearby modes in the spectrum, is their density. These two effects together
determine how much spatial overlap nearby modes in the spectrum typically
have.  A useful quantity reflecting this can be defined by summing the
participation ratios of all modes up to a given point $\lambda$ in the
spectrum. We call this the {\em volume fill fraction} (VFF). It is
essentially the fraction of the total four-volume occupied by modes up to a
given eigenvalue $\lambda$. In Fig.\ \ref{fig:evsvolume24} we show this as a
function of the eigenvalue $\lambda$. In the lowest part of the spectrum where
the spectral density is small and modes are localized, the VFF is much less
than unity allowing for eigenmodes to occur without substantial spatial
overlap. If the eigenmodes are really produced in independent subvolumes the
level statistics is expected to be Poissonian. In contrast, higher up in the
spectrum the VFF becomes much bigger than unity and eigenmodes there must
overlap. Here the eigenvalue statistics is expected to be described by random
matrix theory.

\begin{figure}
\begin{center}
\includegraphics[width=0.6\columnwidth,keepaspectratio]{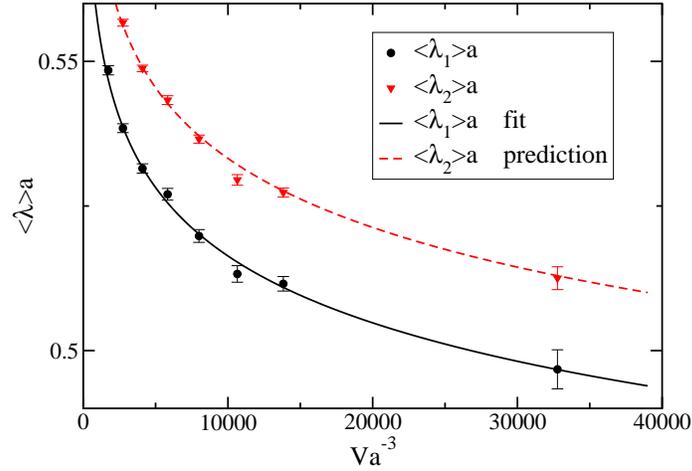}
 \caption{\label{fig:fitsmallest_two} The volume dependence of the smallest
   two eigenvalues of the overlap Dirac operator for $N_t=4$ and Wilson
   $\beta=2.6$. The solid line is a two parameter fit ($\alpha, C$) to the
   analytic prediction based on Poisson statistics and power law spectral
   density. The dashed curve is not a fit, it is the prediction for the second
   smallest eigenvalue based on the already fitted values of $\alpha$ and $C$.
 }
\end{center}
 \vspace{-5mm}
\end{figure}

There are several ways of explicitly checking these expectations on lattice
Dirac spectra. The Poisson and random matrix statistics pertain to two
different universal ways the eigenvalue fluctuations can be correlated locally
in the spectrum. An additional, non-universal feature of the spectrum is the
spectral density $\rho(\lambda)$. In the case of the overlap operator the
spectral density of the low modes is numerically found to be well described by
a simple power law $\rho(\lambda)=C\lambda^\alpha$. Using this and assuming
statistically independent spectral fluctuations for the lowest eigenvalues
(Poisson statistics) we can derive detailed predictions for the distribution
of the lowest first, second, etc.\ eigenvalues and their dependence on the
spatial volume of the lattice. As an illustration, in
Fig.\ \ref{fig:fitsmallest_two} we compare the spatial volume dependence of
the lowest two eigenvalues with the analytic predictions based on Poisson
statistics and the power law spectral density. The solid line is a
two-parameter fit of this analytic form to the average lowest eigenvalue. The
dashed line is a parameter-free prediction for the average of the second
smallest using the already fitted parameters. Both the fit and the
parameter-free prediction describe the data correctly, suggesting that the
smallest two eigenvalues indeed obey Poisson statistics.

For the staggered Dirac operator we did not find a simple analytic description
of the spectral density, but we had much more eigenvalues per
configuration. This allowed us to compute the so called unfolded level spacing
distribution. Unfolding is a standard way in random matrix theory to locally
rescale the spectrum in order to get rid of the dependence on the
non-universal spectral density. If the eigenvalues are statistically
independent, after unfolding the levels are described by a simple
Poisson distribution with exponentially distributed level spacings. If on the
other hand, the original spectrum follows random matrix statistics, the
distribution of the unfolded level spacings is described by the so called
Wigner surmise of the given random matrix ensemble
\cite{Verbaarschot:2000dy}. In the case of the staggered Dirac operator the
corresponding ensemble is the chiral symplectic ensemble.

\begin{figure}
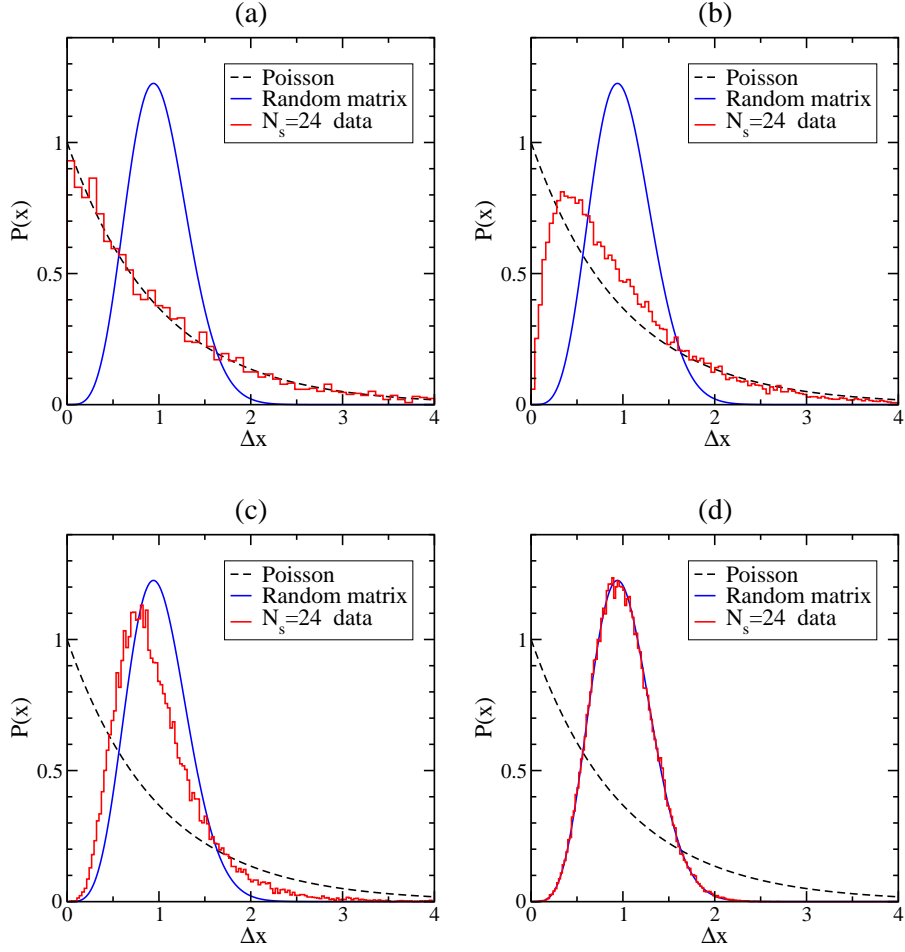

%\vspace{1cm}
\begin{center}
\begin{tabular}{ll}
\includegraphics[width=0.38\columnwidth,
                       keepaspectratio]{ks_levsp_0.00-0.18.eps} &
\includegraphics[width=0.38\columnwidth,
                       keepaspectratio]{ks_levsp_0.22-0.24.eps} \\[5mm]
\includegraphics[width=0.38\columnwidth,
                       keepaspectratio]{ks_levsp_0.245-0.255.eps} &
\includegraphics[width=0.38\columnwidth,
                       keepaspectratio]{ks_levsp_0.28-0.297.eps}
\end{tabular}
\caption{\label{fig:ks_levsp} The panels show the unfolded level spacing
  distribution in different regions of the spectrum. The labeling (a)-(d)
  corresponds to the regions indicated in Fig.\ 2 %\ref{fig:evsvolume24}
  with the shaded areas. The curved lines are the exponential
  distribution and the Wigner surmise.}
\end{center}
 \vspace{-5mm}
\end{figure}

In Fig.\ \ref{fig:ks_levsp} we show the unfolded level spacing distribution of
the eigenvalues in four different regimes of the spectrum. Also shown in each
figure is the universal (parameter-free) prediction based on Poisson and
random matrix statistics respectively. It is apparent that the lowest part of
the spectrum obeys Poisson statistics and that going up in the spectrum this
gradually crosses over to random matrix statistics. 

\begin{figure}
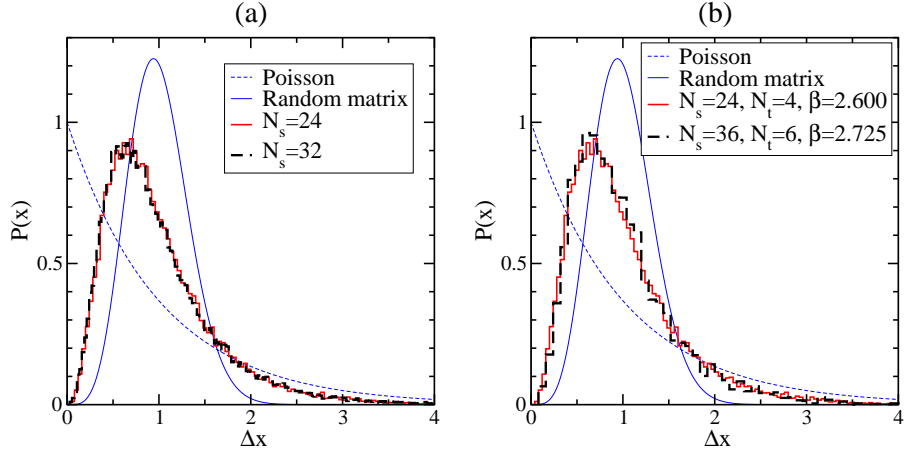

\begin{center}
\begin{tabular}{ll}
\includegraphics[width=0.38\columnwidth,keepaspectratio]{levsp_volscale.eps}
 & 
\includegraphics[width=0.38\columnwidth,keepaspectratio]{levsp_cont.eps}
\end{tabular}
\caption{\label{fig:levsp_scale} The unfolded level spacing distribution for
  two different spatial volumes at fixed $N_t$ (a) and for two different
  $N_t$'s at fixed physical temperature and physical volumes (b). }
\end{center}
   \vspace{-5mm}
\end{figure}

An important question is whether this phenomenon survives the thermodynamic
and the continuum limit. When scaling up the spatial volume one expects that
the number of ``modes of a given statistics'' scales proportionally to the
volume. This can be most simply checked for the eigenvalues of intermediate
statistics. In Fig.\ \ref{fig:levsp_scale}(a) we plotted the level spacing
distribution of eigenvalues 10-20 in a spatial volume of $24^3$ and that of
the scaled up percentage of eigenvalues 24-47 in a volume of $32^3$. The
distributions on the two volumes appear to be the same intermediate
distribution demonstrating that the number of eigenvalues of the same
statistics scales with the volume.

To assess what happens in the continuum limit, in
Fig.\ \ref{fig:levsp_scale}(b) we show level spacing statistics for the same
physical temperature and three-volume with two different lattice spacings
corresponding to $N_t=4$ and $N_t=6$. For both cases we plotted the
statistics based on eigenvalues 10-20. Although the difference in lattice
volumes in lattice units is more than a factor of 3, the two distributions
seem to be identical. This shows that the number of eigenvalues with a given
statistics depends on the physical volume, not on the volume in lattice
units. The lowest, Poisson distributed modes might thus be associated with some
physical objects in the gauge field background, the physical density of which is
constant in the continuum limit.

\begin{thebibliography}{99}

%\cite{Verbaarschot:2000dy}
\bibitem{Verbaarschot:2000dy}
  J.~J.~M.~Verbaarschot and T.~Wettig,
  %``Random matrix theory and chiral symmetry in QCD,''
  Ann.\ Rev.\ Nucl.\ Part.\ Sci.\  {\bf 50}, 343 (2000)
  [arXiv:hep-ph/0003017].
  %%CITATION = ARNUA,50,343;%%

%\cite{Forrester}
\bibitem{Forrester}
 P.~J.~Forrester,
 % ``The spectrum edge of random matrix ensembles''
 Nucl.\ Phys.\ {\bf B402}[FS], 709 (1993)


%\cite{Farchioni:1999ws}
\bibitem{Farchioni:1999ws}
  F.~Farchioni, P.~de Forcrand, I.~Hip, C.~B.~Lang and K.~Splittorff,
  %``Microscopic universality and the chiral phase transition in two flavor
  %QCD,''
  Phys.\ Rev.\  D {\bf 62}, 014503 (2000)
  [arXiv:hep-lat/9912004];
  %%CITATION = PHRVA,D62,014503;%%

%\cite{Damgaard:2000cx}
\bibitem{Damgaard:2000cx}
  P.~H.~Damgaard, U.~M.~Heller, R.~Niclasen and K.~Rummukainen,
  %``Low-lying eigenvalues of the QCD Dirac operator at finite temperature,''
  Nucl.\ Phys.\  B {\bf 583}, 347 (2000)
  [arXiv:hep-lat/0003021].
  %%CITATION = NUPHA,B583,347;%%

%\cite{Pullirsch:1998ke}
\bibitem{Pullirsch:1998ke}
  R.~Pullirsch, K.~Rabitsch, T.~Wettig and H.~Markum,
  %``Evidence for quantum chaos in the plasma phase of {QCD},''
  Phys.\ Lett.\  B {\bf 427}, 119 (1998)
  [arXiv:hep-ph/9803285].
  %%CITATION = PHLTA,B427,119;%%

%\cite{GarciaGarcia:2006gr}
\bibitem{GarciaGarcia:2006gr}
  A.~M.~Garcia-Garcia and J.~C.~Osborn,
  %``Chiral phase transition in lattice QCD as a metal-insulator transition,''
  Phys.\ Rev.\  D {\bf 75}, 034503 (2007)
  [arXiv:hep-lat/0611019].
  %%CITATION = PHRVA,D75,034503;%%

%\cite{Gavai:2008xe}
\bibitem{Gavai:2008xe}
  R.~V.~Gavai, S.~Gupta and R.~Lacaze,
  %``Eigenvalues and Eigenvectors of the Staggered Dirac Operator at Finite
  %Temperature,''
  Phys.\ Rev.\  D {\bf 77}, 114506 (2008)
  [arXiv:0803.0182 [hep-lat]].
  %%CITATION = PHRVA,D77,114506;%%

%\cite{Bilgici:2009tx}
\bibitem{Bilgici:2009tx}
  E.~Bilgici et al.\
  %``Fermionic boundary conditions and the finite temperature transition of
  %QCD,''
  arXiv:0906.3957 [hep-lat].
  %%CITATION = ARXIV:0906.3957;%%

%\cite{Kovacs:2008sc}
\bibitem{Kovacs:2008sc}
  T.~G.~Kovacs,
  %``Gapless Dirac Spectrum at High Temperature,''
  PoS {\bf LATTICE2008}, 198 (2008)
  [arXiv:0810.4763 [hep-lat]].
  %%CITATION = POSCI,LATTICE2008,198;%%

%\cite{Kovacs:2009zj}
\bibitem{Kovacs:2009zj}
  T.~G.~Kovacs,
  %``Absence of correlations in the QCD Dirac spectrum at high temperature,''
  Phys.\ Rev.\ Lett.\  {\bf 104}, 031601 (2010)
  [arXiv:0906.5373 [hep-lat]].
  %%CITATION = PRLTA,104,031601;%%

%\cite{Kovacs:2010wx}
\bibitem{Kovacs:2010wx}
  T.~G.~Kovacs and F.~Pittler,
  %``Anderson Localization in Quark-Gluon Plasma,''
  Phys.\ Rev.\ Lett.\  {\bf 105} (2010) 192001
  [arXiv:1006.1205 [hep-lat]].
  %%CITATION = PRLTA,105,192001;%%

\end{thebibliography}
\end{document}